# The two-component physics in cuprates in the real space and in the momentum representation

**Lev P. Gor'kov**[1,3] **and Gregory B. Teitel'baum**[2]
[1] NHMFL, Florida State University, 1800 E P. Dirac Dr., Tallahassee FL 32310, USA
[2] E.K.Zavoiskii Institute for Technical Physics of the RAS, Sibirskii Trakt 10/7, Kazan 420029, Russia

**Abstract.** Gradual evolution of two phase coexistence between dynamical and static regimes in cuprates is first investigated in the real space by making use of available neutron scattering, NMR and μSR data. Analyses of the Hall effect and the ARPES spectra reveals the presence of two groups of charge carriers in LSCO. The $T$-dependent component is due to the thermal activation of bound electron-hole structures seen near antinodal points in the Brillouin zone, thus introducing the two component physics also for the momentum representation. Interpretation of so-called "van Hove bands" undergoes drastic changes. Importance of the findings for pseudo-gap physics is stressed. Relation to some recent STM and STS results is discussed.

## 1. Introduction

Although the mechanism of high temperature superconductivity (HTSC) in cuprates remains unknown, it is worth mentioning that the superconductivity (SC) itself, beside its $d$-wave character, bears all the major features of SC in more common materials, including the double charge ($2e$) for the SC order parameter, as seen, e.g., from the vortex flux quantization. The most controversial issue concerns the normal state and stems from the fact that the very conductivity in cuprates is the result of an external doping, such, for instance, as doping the Mott insulator, the parent antiferromagnet (AFM) $La_2CuO_4$, with the divalent Sr. Numerous experimental facts demonstrate that the resulting system in no case behaves as an ordinary Fermi liquid (FL).

There currently exist compelling experimental evidences that the high $T_c$- cuprates in the part of the $(T,x)$-phase diagram known as the "pseudogap"(PG) region, are actually inhomogeneous in the real space. More recently, it has been shown [1] that the analysis of the ARPES and the Hall effect data together unequivocally indicates in favor of the two- component physics for these systems also in the momentum representation. The presentation below is the attempt to further comprehend these findings to unify them into a self-consistent picture.

We summarize first the NMR and μSR data from which it follows that the two-phase regions in the bulk, while displaying dynamical character at elevated temperatures, gradually freeze down to form heterogeneous static areas at lower temperatures.

Analysis of ARPES and the Hall coefficient data have led us to the conclusion that the so-called van Hove (vH) singularity at the anti-nodal points has nothing to do with any quasi-particle band structure. Instead, in the vicinity of the anti-nodal "patches" a quasi- periodic (with doubled

3 To whom any correspondence should be addressed.

periodicity) structure appears representing an electron-hole bound state with the binding energy value gradually decreasing with doping, $x$. At $x$~0.2 the change of the regime takes place, corresponding probably to a quantum critical point (QCP)

We postulate that inhomogeneous areas seen in the new generation of the surface-sensitive experiments [2-4], where the heterogeneity itself is static due to low-$T$ pinning at the sample's surface, have the same character as in the bulk. Correspondingly, the presence of such "exciton"-like states suggests the new interpretation for the two energy scales observed in these and other experiments. While the low temperature gap around the nodal regions at the Fermi surface (FS) is justly interpreted as the $d$-wave superconducting (SC) gap, the higher energy scale that persists above $T_c$ and is seen near the (0,π)- points is related to the spatially inhomogeneous distribution of the bound $e$-$h$-pairs and may have nothing in common with a SC gap. We speculate that the aforementioned "excitonic" $e$-$h$-bound states come about from clusters with the predominant participation of the O-Cu pairs and involve the octahedral distortions.

In the discussion below we restrict ourselves mainly by the class of $La_{2-x}Sr_xCuO_4$ (LSCO) materials, for which the most extensive experimental results are available.

## 2. Evolution of the two phase regime with temperature

Experimentally the notion of two components should be dated back to the Johnston's experiments [5] where the decomposition of the magnetic susceptibility of LSCO into the Pauli- and a temperature dependent contributions had been performed for the first time.

Theoretically the tendency to the phase separation in cuprates into magnetic and conducting islands has been predicted in [6] with lattice interactions between neighboring $Cu^{2+}$ ions as the driving mechanism, and, later, in [7] where the phase separation would be due to the exchange interactions between the copper spins. In both cases the electro-neutrality that otherwise would be violated by the presence of immobile Sr-charges was expected to severely restrict sizes of an AFM island, thus resulting in a dynamically frustrated first order transition. As shown in [8], it is energetically preferable for AFM phase, at least at small $x$, to appear in a form of so-called "stripes": charged incommensurate antiferromagnetic structures. The dynamical character of stripes' formation at temperatures above $T_c$ has been demonstrated by observation of large peaks at the incommensurate wave vectors' values around the commensurate AFM (π,π)-point [9] only in the inelastic neutron scattering experiments.

The dynamical spatial separation into two components in *the real space* has been established in [10]. In [10] it was shown that for such local probe experiments as NMR and NQR, the nuclear magnetic relaxation time of the Cu-nuclei spins, $1/{}^{63}T_1(T,x)$, consists of the two contribution:

$$1/{}^{63}T_1(T,x) = 1/{}^{63}\overline{T}_1(x) + 1/{}^{63}T_1(T) \qquad (1)$$

The first component comes about when the participating nuclear spin belongs to a stripe area, while the second one is inherent to the stoichiometric $CuO_2$-plane. This last conclusion follows from the surprising fact that the concentration-independent term, $1/{}^{63}T_1(T)$, turned out to be same for a whole class of materials and, in particular, exactly coincides with the relaxation rate $1/{}^{63}T_1(T)$ measured for the two-chains' material, YBCO 1248 [11].

The standard analysis [10] with the use of the expression:

$$1/T_1 = \frac{k_B T}{2\mu_B^2 \hbar^2 \omega} \sum_i F(Q_i) \int \frac{d^2 q}{(2\pi)^2} \cdot \chi''(q, \omega \to 0) \qquad (2)$$

and of the experimental inelastic neutron scattering data with $\chi''(q,\omega,T)$ from the incommensurate peaks [12] leads to the correct estimate of the first term value in Eq. (1). Even more important, this contribution turns out to be temperature-independent in the whole temperature interval up to 300 K.

Both contributions are, of course, positive, confirming their origination from the two independent relaxation processes.

It is worth mentioning that $1/{}^{63}\overline{T}_1(x)$-value also provides an insight into the intensity of processes related to stripes for a whole class of cuprates. In Fig. 1 from [10] $1/{}^{63}\overline{T}_1(x)$ for LSCO is plotted as a function of the doping, $x$. Although its value rapidly decreases with the doping increase, it always remains considerably larger (by one order of magnitude, approximately) than the same quantity in other cuprates, as shown in the Inset to Fig. 1. In other words, phenomena related to the dynamical stripes' formation are less pronounced in other compounds. Therefore, in what follows we mostly concentrate on the LSCO materials for which the interplay of SC with the magnetic (antiferromagnetic) phenomena seems to be strongest and has been extensively studied.

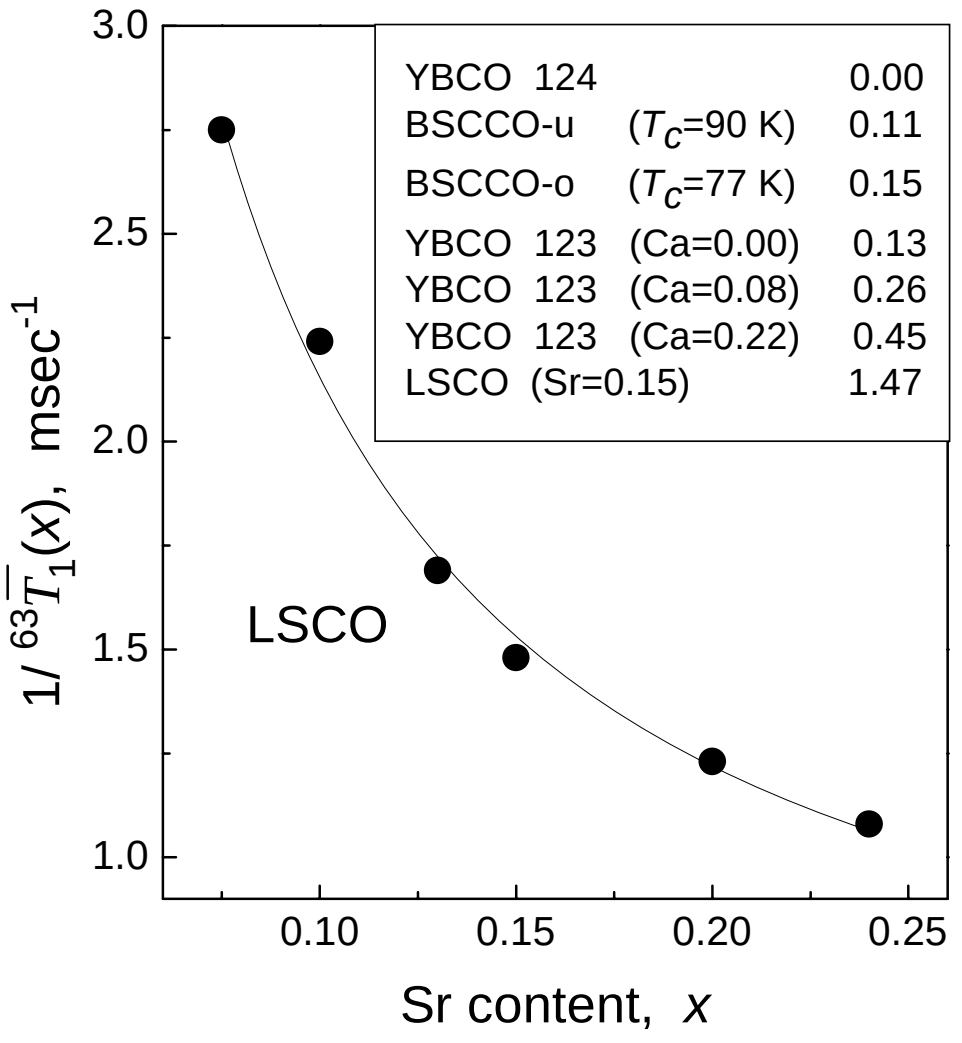

**Figure 1.** The $1/{}^{63}\overline{T}_1(x)$ vs Sr content $x$ for LSCO (together with $1/x$ fit). Inset: $1/{}^{63}\overline{T}_1(x)$ for few other compounds, (u) and (o) indicate underdoped (u) and overdoped (o) BSCCO 2212. Details and references in [10].

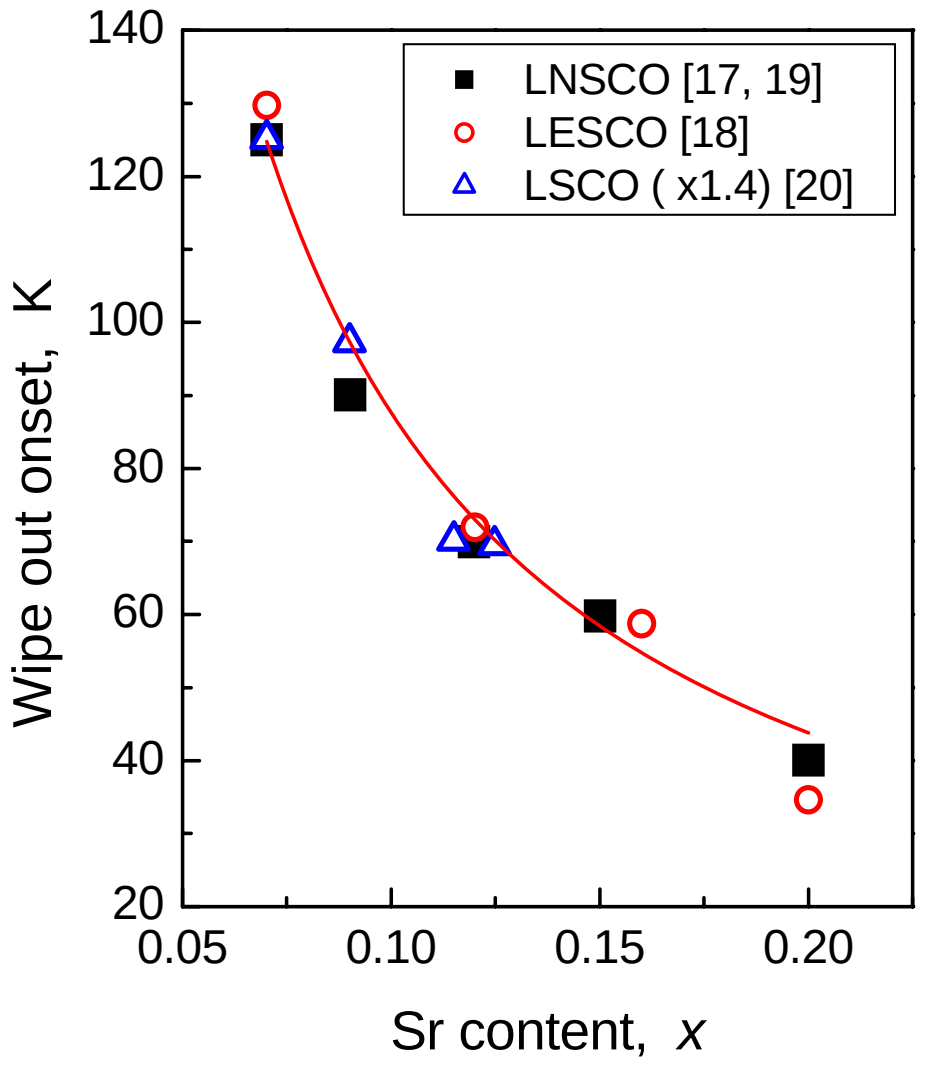

**Figure 2.** Doping dependence of the wipeout onsets for LSCO and rare earth co-doped materials LNSCO (Nd=0.4) and LESCO (Eu=0.2) together with $1/x$ fit. Note, that LSCO data are multiplied by a factor 1.4.

Note that at lower temperatures the coexistence of two static phases is a well documented experimental fact. Indeed, the static stripe phase has been directly seen in the elastic neutron experiments [13] for the Nd-and Eu- doped LSCO (LNSCO and LESCO below). The NQR copper spectra for LESCO at 1.3 K also reveal the in-plane distribution of the hole and spin densities typical of the static stripe phase [14]. For $La_{1.88}Sr_{0.12}CuO_4$ the presence of both SC and the magnetic phases (the volume fraction ~18% for the latter) has also been proven by μSR [15].

At temperatures below the LTT transition for the 1/8-Ba doped LCO (LBCO) the incommensurate antiferromagnetism (ICAFM) manifests itself as the thermodynamic equilibrium state through the magnetic susceptibility anisotropies and the observation of the spin-flop transition [16].

Whether the static magnetic phase occupies the whole sample's volume or only a part of it, can be also judged from measurements of the "wipe-out" fraction of the $^{63}$Cu-signal. This fraction constitutes almost 100% in LNSCO and LESCO [17-19]. Meanwhile, the "wipe-out" is far from being complete in LSCO, thus confirming again the static coexistence of the two different ground states at low temperatures in these materials [20].

On the other hand, as we have said above, the ICAFM-islands seem to manifest themselves at higher temperature only in a dynamical regime. Study of the temperature dependence for the "wipe-out" of the $^{63}$Cu-signal allows to trace in details that the dynamical fluctuations gradually slow down to finally form static patterns at the lowest temperatures [14, 17-20]. In Fig. 2 the temperatures for onset of such a freezing are given for LNSCO, LESCO and LSCO materials. It seems that phases' pinning is governed by the charges of the dopants' ions.

One concludes from this Section that two phases may, indeed, coexist in the sample's volume statically at low temperatures, while at elevated temperatures heterogeneity in the real space manifests itself in a dynamical fashion.

### 3. Two components in the number of charge carriers

Doping of a Mott insulator is not the thermodynamic process, and so far is not understood well. It is commonly accepted that doped holes initially go onto the oxygens' sites and (at $x$>0.02 for LSCO) start to participate in conductivity and other transport phenomena. Naturally, for the above picture of the dynamical coexistence of two phases all transport properties are expected to behave different from predictions of any Boltzmann-like theory. No approach has been developed yet to address the issue theoretically, and we limit ourselves by the mere analysis of experimental data. Fortunately, the detailed data are now available on the temperature dependence both for resistivity and the Hall coefficient in LSCO in the whole range of concentrations and temperatures up to 1000 K [21].

In [1] it has been demonstrated that analyses of the data [21] for the Hall coefficient, $R_{\mathrm{Hall}}(T, x)$, again unambiguously leads to the notion of two components, now in the effective number of carriers:

$$n_{\mathrm{eff}} = n_0(x) + n_1 \cdot \exp(-\Delta(x)/T) \ . \qquad (3)$$

The first component is temperature independent, while the second one bears the activation character. The presence of the activation contribution into Eq. (3) must be considered as the thermodynamic feature of the system.

In Fig. 3 we plot data [21] for $R_{\mathrm{Hall}}(T, x)$ against Eq. (3) to demonstrate the excellent fit for the whole concentration range (above temperatures where localization features begin to be seen in the resistivity behavior).

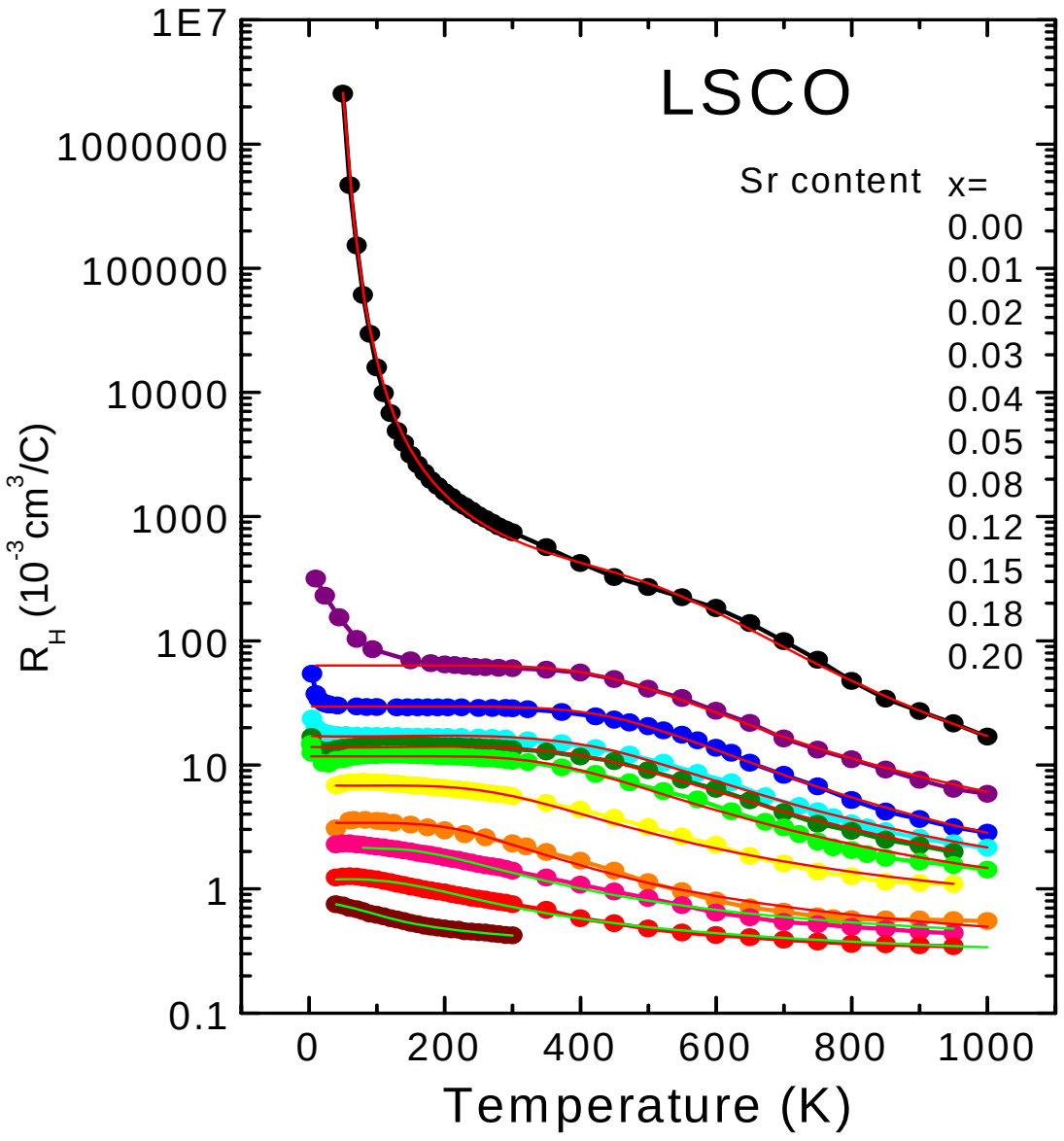

**Figure 4.** The doping dependent evolution of $R_{\mathrm{H}}$ [21] with fittings to Eq. (3).

Dependence of $n_0(x)$ presented in [1] does show that already at $x$=0.07 the number of carriers at low temperatures, $n_0(x)$, starts to exceed the number of the externally doped holes and rapidly increases with $x$ to values close to one carrier per Cu-site. Similar low temperature doping behavior for the Hall coefficient has also been observed in the single–layer BSLCO [22]. We emphasize that our analysis in terms of Eq. (3) was restricted by the temperatures above 100 K. At low temperatures $R_H$ is not monotonous for $0.15 \le x \le 0.17$. This may be due to interplay of holes with the charge carriers from the electron pockets whose mobility is increasing upon cooling [23]. The observed picture may be also affected by the stripes' pinning and by an increase in hybridization between the oxygen- and the copper-orbitals.

The coefficient $n_1(x)$ in front of the exponent is practically concentration-independent ($n_1(x)$ ~2.8), but rather abruptly drops to zero at $x$ above ~0.2.

It is crucial to emphasize once more the activation character of the $T$-dependent contribution. While in [24] the decrease of $R_{Hall}(T, x)$ with temperature has been ascribed to a "spill-over" of electrons from the so-called vH "flat bands" near the antinodal points (0,π), such an interpretation would be unable to explain the activation form of the $T$-dependence in Eq. (3). The vH-type singularity with its logarithmic character, it is too week to account for the exponential term in (3). For the activation gap in Eq.(3) to exist there must be a sharp feature in the spectrum of the system, with a large density of states that is well separated from the position of the chemical potential. (Note also the large value of the pre-factor, $n_1$ ~2.8 proportional to the density of states (DOS) per unit cell).

Eq. (3) also adequately describes the earlier data for $R_{Hall}(T, x)$ for temperatures and compositions available in [25] and [26]. (The $x$-dependence of the gap, $\Delta(x)$, itself is discussed in the next sub-Section).

## 4. Comparison with the ARPES results

The activation term itself in Eq. (3) of the previous Section could, in principle, be interpreted in sort of a band picture as if due to the thermal excitation of an electron from the chemical potential to the bottom of an unoccupied narrow band or an empty energy level. In [21] $R_{Hall}(T, x)$-data are decomposed into a "nodal" term, $R_{Hall}(T, x)^{arc}$ , and $R_{Hall}(T, x)^{LFS}$, a contribution from "the large FS". For the latter the activation gap is taken as the $x$-independent "charge transfer" gap (from its value at small $x$), while additional $T$-dependences for the two components come about through $T$-dependence of their relative fractions. Both assumptions [21], however, are not consistent with the ARPES data [27-29].

In the ARPES experiments one measures the energy of out-coming photoelectron with a known momentum with respect to the chemical potential's energy position. (Theoretically, ARPES can "reconstruct" the electronic band spectra of a system in the whole Brillouin Zone (BZ) provided electrons were not interacting strongly). In [27-29] data are available, in particular, for the quasi momentum near the vicinity of the antinodal (vH) points, (0,π). Ideally, according to the band picture, the electronic ARPES spectra at these points would determine the (negative) energy of the "vH flat band" with respect to the chemical potential. These data are now shown in Fig. 4 together with the values for $\Delta(x)$ extracted from the $R_{Hall}(T, x)$ [21]. The two quantities *coincide* within the limits of experimental errors. (In Fig. 4 new points added [29] to Fig. 2 of [1]; the points shown by green squares are discussed later).

This is the surprising result. Indeed, it is difficult to believe that the thermal excitation of an electron from the chemical potential to a hypothetical energy level that lies well *above* the latter (to create a hole, as needed for Eq. (3)), would cost practically the same energy as to transfer to the chemical potential an electron taken from the "vH flat band" that lies well *below* it (at least at small $x$). Although a numerical coincidence may occur at some doping, Fig. 4, as a whole, obviously, manifests the new physics.

In [1] we made a comment that the result is an indication in favor of localized states near the antinodal points. Here we elaborate further that the equal energy for creation of an electron (ARPES) and a hole (thermal activation) can be understood only in terms of *bound* states for electron-hole pairs. These states, being positioned in the momentum representation near the vH points, according to ARPES, form quasi-periodic structures (close to the double periodicity along the bonds directions). Large DOS (large $n_1$~2.8 in Eq. (3)!) show that these formation densely populate the real space.

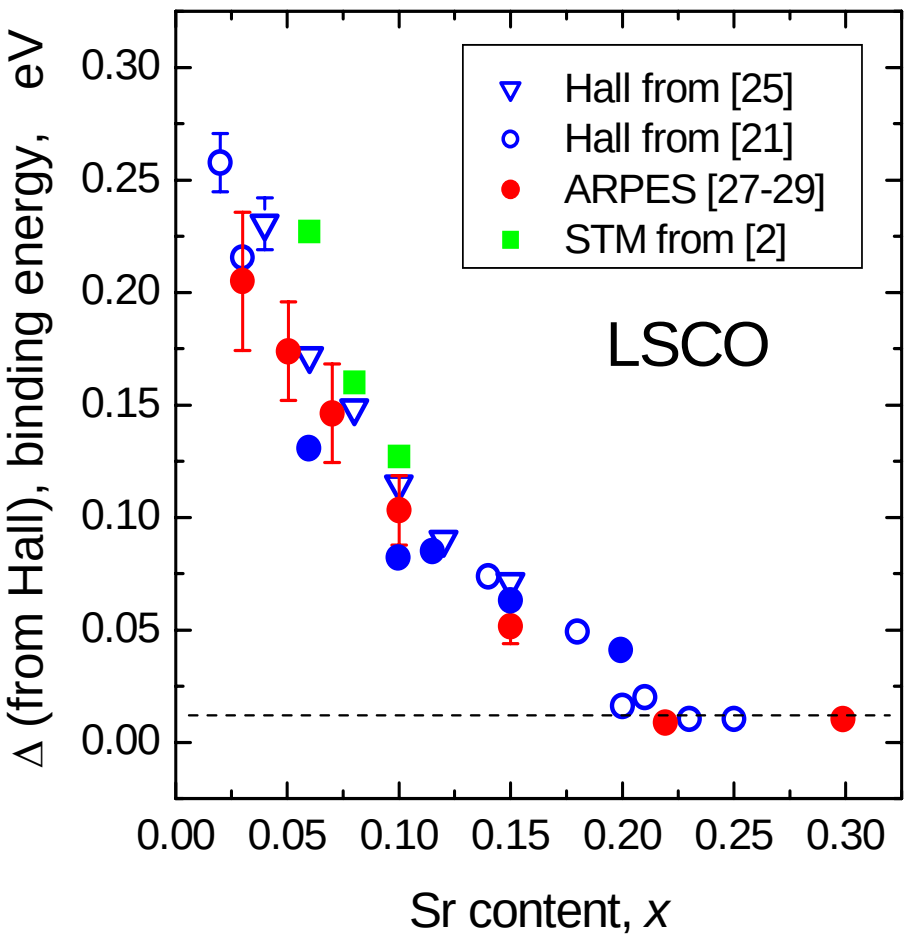

**Figure 4.** The ARPES binding energy vs the Hall activation energy. (See [1] for the details.)

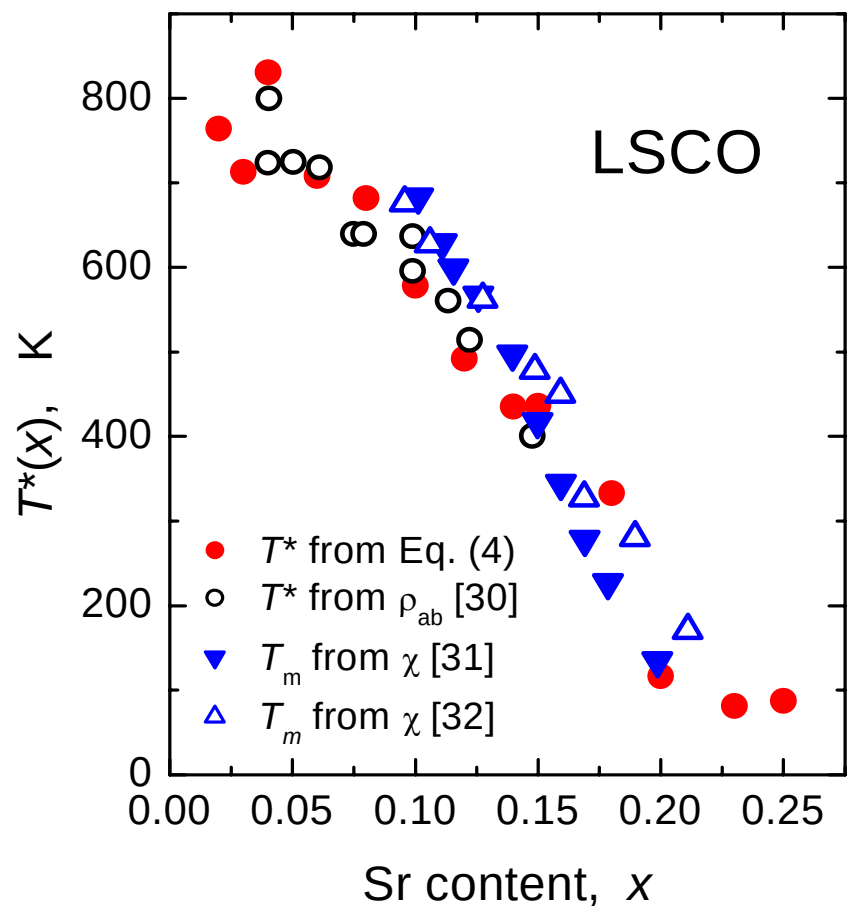

**Figure 5.** The doping dependence of $T^*(x)$ for $La_{2-x}Sr_xCuO_4$ [1].

ARPES is the fast experiment. Each emitted photoelectron leaves behind the perturbed lattice. In other words, the energy cost, as seen by ARPES, for breaking the bound *e-h* pair should be less than the energy needed for breaking the same bound pair in the thermal activation process. This feature is indeed present in Fig. 4: all points from the ARPES data [27-29] lie below the corresponding $\Delta(x)$-values by some ~ 15 meV, a scale more typical for the lattice effects.

In Fig. 4 one can notice some sharp feature at $x$ above ~0.2, the same concentration at which $n_1$ begins to drop to zero [1]. However, the binding energy does not go to zero: from more recent ARPES data [29] one concludes that the bound state persists above $x$~0.2, as shown by the dashed line. The energy gap still persists, but now is of order of ~ 15 meV. We speculate that the lattice effects are also involved into formation of such *e-h* structures.

Change with temperature of the balance between number of carriers introduced through external doping process, $x$, and the concentration of the thermally activated holes should strongly impact transport and other characteristics of cuprates. In Fig. 5 we plotted the temperature at which the number of carriers from the each two components becomes equal:

$$T^*(x) \approx -\Delta(x) / \ln x \qquad (4)$$

Other symbols in Fig. 5 show the marks derived from behavior of the magnetic susceptibility and resistivity data at their crossing into the pseudogap regime in LSCO. The result shows that Eq. (4) actually determines so-called pseudogap temperature, $T^*(x)$, quite well. With the increase of number of carriers at elevated temperatures to ~ one carrier per Cu one may expect the restoration of the FL features in cuprates, including a large Fermi surface.

## 5. Brief discussion

We do not attempt to provide any strict theoretical interpretation. The facts above are phenomenological ones, obtained from the analyses of experimental data, and below we shall suggest a few speculations only.

To start with, one may ask whether same energy scales, as shown in Fig. 4 for LSCO, have ever been seen in other materials, by other methods. APRES data, while abundant for two-layer systems like YBCO or BSCO 2212, are unable to provide a reliable answer being masked by the inter-plane bonding / anti-bonding splitting of spectra near the "vH patches".

In Fig. 6 we reproduce STM results from [2]. Positions of the ticks on the right side of *dJ/dV* curves are plotted as the green squares in Fig. 4 for three concentrations. The agreement with the data for LSCO is excellent. More intriguing is that, according to the STM map in Fig. 6, these features are seen for *dJ/dV* –curves in the bright ("*metallic*") regions. It is reasonable to assume that the shoulders indicated by ticks correspond to the bias that matches the "ionization" energy of the bound *e-h* states discussed above. Although in [2] resolution is worse than the atomic scale, the pattern of bright and dark pinned spots shown there, looks very similar to the one in Fig. 3 of [3] for the two materials, with the Dy-doped BSCO 2212 among them.

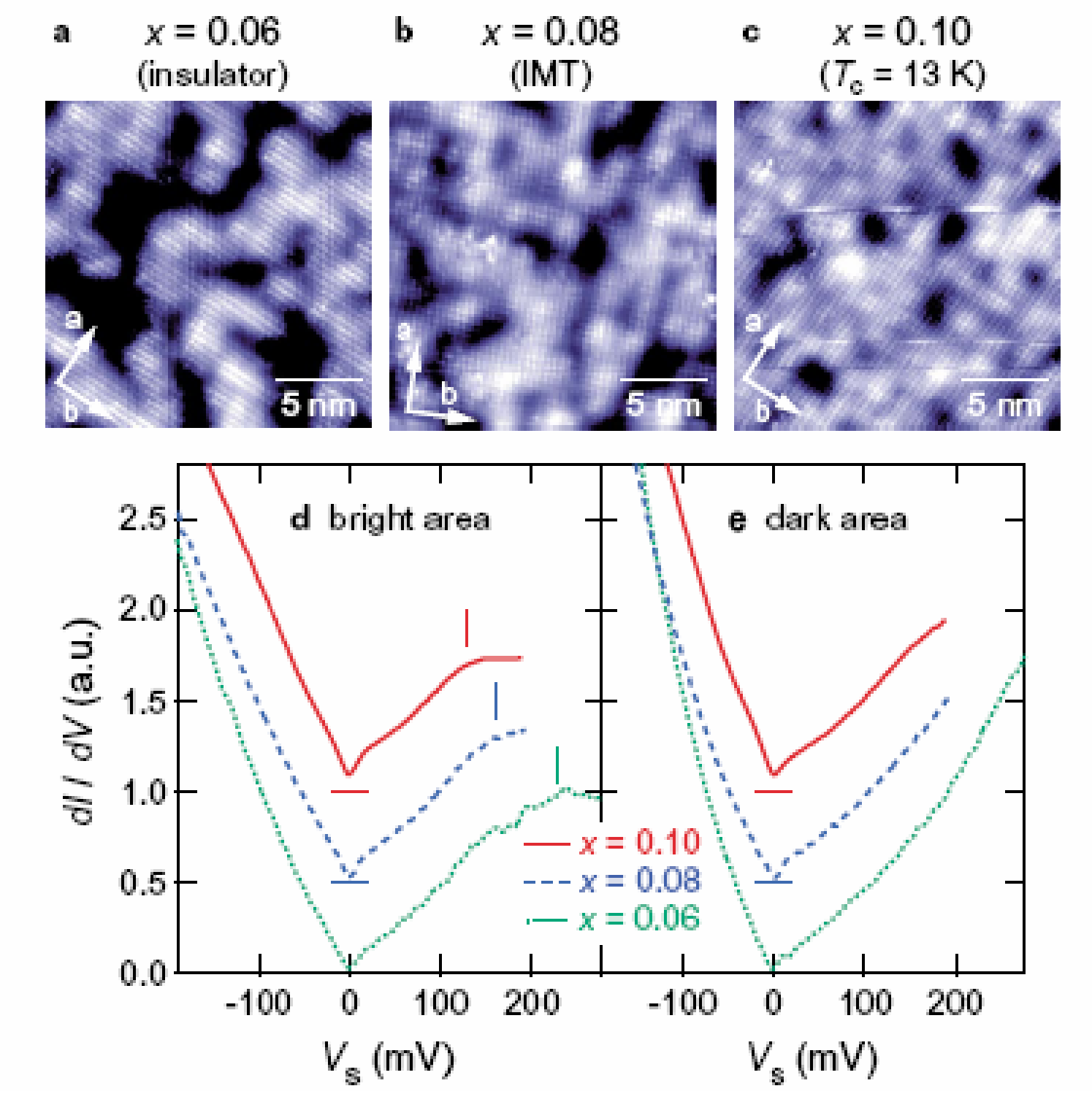

**Figure 6.** *dJ/dV* curves obtained by STM imaging/spectroscopy of $Ca_{2-x}Na_xCuO_2Cl_2$ [2]. Positions of the ticks on the right side of *dJ/dV* curves are plotted as the green squares in Fig. 4 for three concentrations.

It should be supposed that the "vH" spots in the BZ observed within the ARPES accuracy are in some correspondence with the inhomogeneous distribution of pinned nanoscale clusters seen in STS by the Davis group and others (See [3] and references there in). From the results above it becomes clear that the energy values plotted in Fig. 4 are actually related to the inhomogeneous electronic structure. Therefore, it seems that the language of the two "gaps", often used in the literature (see, e.g., [33]), i.e., of the SC gap and so-called "pseudogap" that would persist above *Tc* near (0,π) points, can only obscure the real physics. There is no "pseudogap": what one sees near the antinodal, or "vH", points corresponds to the internal scale of the bound electron-hole structures in Fig. 4. Their double-periodicity is heterogeneously distorted along the sample at low temperatures due to random pinning.

To emphasize again, no theoretical description exists for the *e-h* bound states that we hypothesized above. One can only speculate that adjacent pairs of oxygen and copper ions may be involved, and the hybridization between O- and Cu-orbitals results in an intermediate valence of the Cu-ions belonging to a cluster. Unlike the Zhang-Rice singlet, such a construction may possess a spin. One may anticipate a local variation of distances between O and Cu. The current STM or STS experiments are not sensitive enough so far to verify such an assumption.

**6. Acknowledgements**

The work of L.P.G. was supported by the NHMFL through NSF cooperative agreement DMR-9527035 and the State of Florida, that of G.B.T. through the RFBR Grant N 07-02-01184.